\documentclass[showpacs,groupedaddress,nofootinbib,superscriptaddress,10pt]{revtex4}
\usepackage{amssymb}
\usepackage{graphicx}
\usepackage{dcolumn}
\usepackage{bm}
\usepackage{graphicx}
\usepackage{epsfig}
\usepackage{color}
\usepackage{graphics}
\usepackage{amsthm}
\usepackage{amsthm}
\usepackage{dcolumn}
\usepackage{float}
\usepackage{ulem}

\begin{document}

\title{Interacting cosmic fluids and phase transitions under\\
a holographic modeling for dark energy}
\author{Samuel Lepe}
\email{samuel.lepe@pucv.cl}
\affiliation{Instituto de F\'\i sica, Facultad de Ciencias, Pontificia Universidad
Cat\'olica de Valpara\'\i so, Avenida Brasil 4950, Valpara\'\i so, Chile.\\
}
\author{Francisco Pe\~{n}a}
\email{francisco.pena@ufrontera.cl}
\affiliation{Departamento de Ciencias F\'isicas, Facultad de Ingenier\'ia y Ciencias,
Universidad de La Frontera, Casilla 54-D, Temuco, Chile.}
\date{\today }

\begin{abstract}
We discuss the consequences of possible sign changes of the $Q$-function
which measures the energy transference between dark energy and dark matter.
We investigate this scenario from a holographic perspective to modeling the
dark energy by a linear-parametrization of the equation of state parameter
denoted by $\omega$. By imposing the strong constraint of the second law of
thermodynamics, we show that sign changes of $Q$ due to the cosmic evolution
imply changes in the temperatures of dark energy and dark matter, respectively.
We also discuss the phase transitions, in the past and future, experienced by dark
energy and dark matter (or, equivalently, the sign changes of their heat
capacities).
\end{abstract}

\maketitle

\section{Introduction}

Certain observational evidence seems to indicate that dark energy
and dark matter, considered as dominant components of the cosmic
fluid, can interact \cite{Abdalla2014} or perhaps an interaction
between dynamical vacuum energy and matter it is also possible
\cite{Basilakos}. Appears unnatural to think that cosmic fluids
coexist and do not interact with each other as is usual in the
standard $\Lambda CDM$ model (both components evolve independently
and satisfy usual energy conservation equations). As we know,
$\Lambda CDM$, despite their successes, is theoretically
unappealing because of several well known shortcomings, among
other, a negative energy density for the dark energy at $z=2.34$
\cite{Delubac} and the case $\omega <-1$ (phantom dark energy,
which is not ruled out by the observational data), an interesting
issue which it seems very difficult to understand within the
$\Lambda CDM$-framework.

On the other hand, dark energy and dark matter, both, seen as ordinary fluids
(perfect fluids), is a simple assumption which is consistent to describe
them. In other words, here, dark energy and dark matter are described under a
representation of perfect fluids and in the literature we see nothing
against this assumption. Any case, it is an interesting approach if we are
thinking that a perfect fluid is something that we know how to handle.

Components under interaction lead to a new perspective for
describing/visualizing the cosmic evolution. There is a rich
literature on the subject and for describing that interaction, for
instance, various Ansatzes for $Q$ (defined before) are used. In
other words, the $Q$ function is put by hand given that the field
equations so require. That is a first approximation to describe
the interaction required by the observational data, but there is
not formalism that allow us to obtain $Q$ from first principles.

In this work, we analyze thermodynamical aspects from the aforementioned
interaction in the framework of two interacting fluids. And as already said,
we will use the holographic philosophy for modeling the dark energy where
the emphasis will put on the temperature evolution, during the cosmic
evolution, from each fluid (dark energy and dark matter. Phase transitions
(seen as sign changes in the heat capacity) will be also discussed. $8\pi G=$
$c=1$ units will be used.

The organization of the paper is as follows: in Sec. II we present a brief
description of two non-interacting fluids and we revise the behaviour of its
temperatures through the cosmic evolution. In Section III we incorporate
interaction and we revise the involved thermodynamics bearing in mind the
second law. In Section IV we we discuss the interaction under holographic
considerations and we show the presence of phase transitions (sign changes
in its heat capacities) experienced by both fluids during the evolution.

Finally, Sec. V is devoted to conclusions. We have added an Appendix in
order to show explicitly the sign changes in the heat capacities of the
interacting fluids.

\section{Non-Interacting fluids and Thermodynamics}

We consider, in the framework of General Relativity (GR), the
non-interacting flat-FLRW scheme between two components
\begin{equation}
3H^{2}=\rho _{de}+\rho _{dm},
\label{eq:eq1}
\end{equation}
and
\begin{eqnarray}
\dot{\rho}_{de}+3H(1+\omega _{de})\rho _{de}&=&0, \nonumber \\
\dot{\rho}_{dm}+3H(1+\omega _{dm})\rho _{dm}&=&0,
\label{eq:eq2}
\end{eqnarray}
where $\rho _{de}$ denotes the dark energy density, $\rho _{dm}$
denotes the dark matter density, $H$ is the Hubble parameter and
dot denotes derivative with respect to the cosmic time. We assume
that both components can be amenable to study by using a
description under the rigorous scope of the thermodynamic laws.
Then, by using the equation $TdS=d\left( \rho V\right) +p\;dV$,
for a generic fluid, where $T$ is the temperature, $S$ the
entropy, $\rho $ the energy density, $p$ its pressure and $V$ the
physical volume, besides the integrability condition $\partial
^{2}S/\partial T\partial V=\partial ^{2}S/\partial V\partial T$,
we can obtain the following equation for the temperature during
the cosmic evolution \cite
{Maartens1996,Zimdahl1997,RadicellaPavon2012}
\begin{eqnarray}
\frac{dT}{T} &=&-\frac{dV}{V}\left( \frac{\partial p}{\partial \rho }\right)
_{V}=-3Hdt\left( \frac{\partial p}{\partial \rho }\right) _{a},  \nonumber
\label{eq:eq3} \\
&\rightarrow &T\left( z\right) =T\left( 0\right) \exp \left( 3\int
\frac{dz}{ 1+z}\omega \left( z\right) \right) ,
\end{eqnarray}
we have used $p=\omega \rho $ , $1+z=a_{0}/a$ with $z$ being the
redshift parameter and $a$ the cosmic scale factor with
$a_{0}=a(0)$. Since $V=\Omega _{0}a^{3}$ (for a spatially flat
section $\Omega _{0}=4\pi /3$, see \cite {ZhangBak2000}), we
obtain $dV/V=3da/a=3Hdt$. By considering for dark energy $\omega
_{de}\left( z\right) \approx -1$ and for non-relativistic dark
matter $p_{dm}=nT_{dm}$ and $\rho _{dm}=nm+3nT/2$, where$T_{dm}\ll
m$ \cite{Maartens1996}, we can obtain $T_{dm}\left( z\right) \sim
(1+z)^{2}$ after using (3). So,
\begin{equation}
T_{de}(z)=T_{de}(0)(1+z)^{-3}\text{ \ },\text{ \ }
T_{dm}(z)=T_{dm}(0)(1+z)^{2},
\end{equation}
i.e., $T_{de}\left( z\right) $ grows and $T_{dm}\left( z\right) $ decreases
when $z$ goes to the future. This fact appears to be unusual if we want to
think in energy transference from dark energy to dark matter, at least from
some $z_{e}$ in the past, and $Q\left( 0\right) >0$ as is suggested by
the observational data \cite{Abdalla2014}. This point will be discussed later.
Equilibrium?\textbf{\ } We do
\[
T_{de}\left( z_{e}\right) =T_{dm}\left( z_{e}\right) ,
\]
\[
\Longrightarrow \;z_{e}=\left[ T_{de}\left( 0\right) /T_{dm}\left(
0\right) \right] ^{1/5}-1,
\]
\begin{equation}
\rightarrow \left\{
\begin{array}{c}
z_{e}\text{ }\epsilon \text{ }\mbox{past}\Longrightarrow T_{de}\left( 0\right)
/T_{dm}\left( 0\right) >1, \\
z_{e}\text{ }\epsilon \text{ }\mbox{future}\Longrightarrow
T_{de}\left( 0\right) /T_{dm}\left( 0\right) <1.
\end{array}
\right.
\label{eq:eq5}
\end{equation}
Nowdays we would expect that $T_{de}\left( 0\right) >T_{dm}\left(
0\right) $ (see later) and in this case $z_{e}$ belongs to the
past. But the problem is \textbf{\ }$z_{e}$\textbf{\ }at the
future. If we do not have equilibrium through the evolution, we
can think in two options: only one sign of $Q$ and then only one
option for the inequality between the temperatures, $T_{de}\left(
z\right) >T_{dm}\left( z\right) $ or $T_{de}\left( z\right)
<T_{dm}\left( z\right) $. Nevertheless, there are holographic
arguments in order to justify, at least in a theoretical scope,
both options for the sign of $Q$ \cite{ArevaloCLP2014}\thinspace .
As a second option we can imagine that even when there are sign
changes in $Q$, the amount of energy transferred is not enough for
generating changes in the relation between $ T_{de}\left( z\right)
$ and $T_{dm}\left( z\right) $. But the thermodynamical constraint
given by the second law rejects the latter option.

By following the reference \cite{Maartens1996}, we  ascribe a
temperature (Gibbs integrability condition) to the dark energy in
the form $T_{de}\sim \rho _{de}^{\omega / \left( \omega +1\right)
}$, and we see that if $-1<\omega <0$, $T_{de}$ increases when
$\rho _{de}$ decreases and vice versa. If we assume that the given
relationship between temperature and energy density is valid for
$\omega \rightarrow -1$ (phantom dark energy, not ruled out by the
observational data) we have that $T_{de}$ and $\rho _{de}$, both
increase with the time and dark energy increasing in the future
appears to be not only a conjeture.

On other hand, if we write for dark matter $T_{dm}\sim \rho
_{dm}^{\omega / \left( \omega +1\right) }$, we can see that
$\omega =0$ (dust) drives to $T_{dm}=const$ and if this were the
case, we would have, from a thermodynamical point of view, a fluid
without the ability of interacting with other and, in this sense,
we could accept the idea developed in the reference \cite{Gomez}:
there is interaction, but $Q=0 $. So, if we accept that both
temperatures $T_{de}\left( \omega =-1\right) =0 $ and $T_{dm}=0$,
we repeat, the above mentioned idea appears to be
interesting.\newline

In the next Sections we introduce the interaction and discuss its
consequences for the evolution of the temperatures.

\section{INTERACTING FLUIDS AND THERMODYNAMICS}

By using the interacting scheme
\begin{eqnarray}
\dot{\rho}_{de}+3H\left( 1+\omega _{de}\right) \rho _{de}&=&-Q,\nonumber \\
\dot{\rho}_{dm}+3H\left( 1+\omega _{dm}\right) \rho _{dm}&=&Q,
\label{eq:eq6}
\end{eqnarray}
or, equivalently
\begin{eqnarray}
\dot{\rho}_{de}+3H\left( 1+\omega _{de}^{eff}\right) \rho _{de}&=&0,\nonumber \\
\dot{\rho}_{dm}+3H\left( 1+\omega _{dm}^{eff}\right) \rho
_{dm}&=&0,
\end{eqnarray}
where
\begin{equation}
\omega _{de}^{eff}=\omega _{de}+\frac{Q}{3H\rho _{de}}\text{ \ \ }\mbox{and}\text{
\ \ }\omega _{dm}^{eff}=\omega _{dm}-\frac{Q}{3H\rho _{dm}},
\label{eq:eq8}
\end{equation}
the temperatures are, respectively,
\begin{eqnarray}
T_{de}\left( z\right)  &\sim &\exp \left( 3\int \frac{dz}{1+z}\omega
_{de}^{eff}\right) \text{ }  \nonumber  \label{eq9} \\
&=&\exp \left( 3\int \frac{dz}{1+z}\omega _{de}\right) \exp \left(
3\int \frac{dz}{1+z}\frac{Q}{H\rho _{de}}\right) , \label{eq:eq9}
\end{eqnarray}
and
\begin{eqnarray}
T_{dm}\left( z\right)  &\sim &\exp \left( 3\int \frac{dz}{1+z}\omega
_{dm}^{eff}\right)   \nonumber  \label{eq10} \\
&=&\exp \left( 3\int \frac{dz}{1+z}\omega _{dm}\right) \exp \left( -3\int
\frac{dz}{1+z}\frac{Q}{H\rho _{dm}}\right) ,
\label{eq:eq10}
\end{eqnarray}
and here we have considered a generic $\omega _{dm}$ for dark matter without
any special commitment with it. In Section IV we will use $\omega _{dm}=0$
(dust).

On the other hand, in presence of interaction, the entropy production
associated to the interaction is
\begin{equation}
\dot{S}_{de}\left( t\right) +\dot{S}_{dm}\left( t\right) =\left(
\frac{ Q\left( t\right) }{T_{dm}\left( t\right) }+\frac{\left(
-Q\left( t\right) \right) }{T_{de}\left( t\right) }\right)
V\geqslant 0,
\end{equation}
$\forall\; t$, or, by using the redshift parameter
\begin{equation}
-\left( 1+z\right) H\left( z\right) \frac{d}{dz}\left[
S_{de}\left( z\right) +S_{dm}\left( z\right) \right] =\left(
\frac{1}{T_{dm}\left( z\right) }-\frac{1}{T_{de}\left( z\right) }
\right) VQ\left( z\right) \geqslant 0,
\end{equation}
$\forall\; z$, i.e. $T_{de}\left( z\right) >T_{dm}\left( z\right)
$ at late times $\Longrightarrow Q\left( z\right) >0$. However, at
early times $ T_{de}\left( z\right) <T_{dm}\left( z\right) $ (a
reasonable assumption), then we should have $Q\left( z\right) <0$
in order to satisfy the second law. As was said before, it may
seem strange that today $T_{de}\left( z\right) $ grows with $z$ if
dark energy is transferring energy to dark matter ($Q\left(
z\right) >0$) at least from some $z_{e}$ (\ref{eq:eq5}). Can be a
signal of a negative heat capacity for the dark energy? As we will
see in the next Section, the incorporation of interaction shows
explicitly this fact. Moreover, the dark matter shows also phase
transitions, in the past and in the future. Sign changes of $Q$
(one in the past and another in the future) can be visualized in
reference \cite {ArevaloCLP2014} where a holographic modeling for
the dark energy was used beside the linear-parametrization $\omega
_{de}\left( z\right) =\omega _{de}\left( 0\right) +\sigma \ast z$,
and $\sigma =\mbox{const}.$\cite{Virey2004}.

We end this Section by making a consistency check. By using the
usual concepts of thermodynamics, we deduce the equation for the
evolution of the temperature for a generic fluid denoted by $\rho
$. We start by setting $ \rho =\rho \left( V,T\right) $ and
$p=p\left( V,T\right) $, a reasonable setup. So,
\begin{equation}
\dot{\rho}=aH\left( \frac{\partial \rho }{\partial a}\right) _{T}+\left(
\frac{\partial \rho }{\partial T}\right) _{a}\dot{T},
\label{eq:eq13}
\end{equation}
and $V=\left( 4\pi /3\right) a^{3}$ so that $\dot{V}=3VH$ and $\left(
\partial \rho /\partial V\right) _{T}=\left( a/3V\right) \left( \partial
\rho /\partial a\right) _{T}$. On the other hand, from the second law
besides the integrability condition given before, it is straightforward to
obtain the expression
\begin{equation}
\left( \frac{\partial \rho }{\partial a}\right) _{T}=\frac{3T}{a}\left[
\left( \frac{\partial p}{\partial T}\right) _{a}-\frac{p+\rho }{T}\right] .
\end{equation}
By replacing this last expression in (\ref{eq:eq13}) and by using the
non-conservation equation
\begin{equation}
\dot{\rho}=Q_{f}-3H\left( \rho +p\right) ,
\end{equation}
where $Q_{f}=-Q$ for dark energy and $Q_{f}=Q$ for dark matter, we can
obtain
\begin{equation}
\frac{\dot{T}}{T}=-3H\left( \frac{\partial p}{\partial \rho
}\right) _{a}+\left( \frac{\partial \rho }{\partial T}\right)
_{a}^{-1}\frac{Q_{f}}{T} ,
\end{equation}
and by using the redshift parameter, we write the last equation in the form
\begin{equation}
\frac{dT}{T}=\frac{dz}{1+z}\left[ 3\left( \frac{\partial
p}{\partial \rho } \right) _{z}-\frac{Q_{f}}{H}\left(
\frac{\partial \rho }{\partial T}\right)
_{z}^{-1}\frac{1}{T}\right] ,
\end{equation}
so that, for dark energy
\begin{equation}
\frac{dT_{de}}{T_{de}}=\frac{dz}{1+z}\left[ 3\left( \frac{\partial
p_{de}}{
\partial \rho _{de}}\right) _{z}+\frac{Q}{H}\left( \frac{\partial \rho _{de}
}{\partial T_{de}}\right) _{z}^{-1}\frac{1}{T_{de}}\right] ,
\label{eq:eq18}
\end{equation}
and for dark matter
\begin{equation}
\frac{dT_{dm}}{T_{dm}}=\frac{dz}{1+z}\left[ 3\left( \frac{\partial
p_{dm}}{
\partial \rho _{dm}}\right) _{z}-\frac{Q}{H}\left( \frac{\partial \rho _{dm}
}{\partial T_{dm}}\right) _{z}^{-1}\frac{1}{T_{dm}}\right] .
\label{eq:eq19}
\end{equation}
Now, we compare the expressions given in (\ref{eq:eq9}) and (\ref{eq:eq18}), i.
e.,
\begin{equation}
\frac{dT_{de}}{T_{de}}=\frac{dz}{1+z}\left[ 3\omega _{de}\left(
z\right) + \frac{Q}{H}\frac{1}{\rho _{de}}\right] ,
\label{eq:eq20}
\end{equation}
and
\begin{equation}
\frac{dT_{de}}{T_{de}}=\frac{dz}{1+z}\left[ 3\omega _{de}\left(
z\right) + \frac{Q}{H}\left( \frac{\partial \rho _{de}}{\partial
T_{de}}\right) _{z}^{-1}\frac{1}{T_{de}}\right] , \label{eq:eq21}
\end{equation}
where we have done $\left( \partial p_{de}/\partial \rho
_{de}\right) _{z}=$ \ $\omega _{de}\left( z\right) $. The
consistency between (\ref{eq:eq20}) and ( \ref{eq:eq21}) indicates
that
\begin{equation}
\rho _{de}=\left( \frac{\partial \rho _{de}}{\partial T_{de}}\right)
_{z}T_{de},
\end{equation}
and the same occurs if we compare (\ref{eq:eq10}) and (\ref{eq:eq19})
\begin{equation}
\rho _{dm}=\left( \frac{\partial \rho _{dm}}{\partial T_{dm}}\right)
_{z}T_{dm},
\end{equation}
as it is expected (see \cite{BrevickNOV2004}).

\section{SIGN CHANGE OF Q AND HOLOGRAPHY}

One approach for treating the $Q$-function is by considering the following
Ansatz
\begin{equation}  \label{eq24}
Q=3H\left( \lambda _{1}\rho _{de}+\lambda _{2}\rho _{dm}\right) ,
\label{eq:eq24}
\end{equation}
where $\lambda _{1}$ and $\lambda _{2}$ are both constant
parameters to be determined by observations. According to
observational settings, both parameters have equal sign and so
there is not sign change in $Q$ \cite {HeWA2009}.

The second approach, which we will use from now on, is based on a
holographic model
\begin{equation}
\rho _{de}\left( z\right) =3H^{2}\left( z\right) \left[ \alpha -\frac{\beta
}{2}\left( 1+z\right) \frac{d\ln H^{2}\left( z\right) }{dz}\right] ,
\label{eq:eq25}
\end{equation}
where $\alpha $ and $\beta $ are both positive parameters which
are well confined by the observational data: $\beta <\alpha <1$
\cite{ArevaloCLP2014, GrandaOliveros2008}.

The infrared cut-off given for $\rho _{de}$
\cite{GrandaOliveros2008} can be understood as a generalization of
the model $\rho _{de}\sim -R$ \cite{Changjun}, where $R$ is the
Ricci scalar given by $R=-6\left( 2H^{2}+\dot{H}+k/a^{2}\right) $
or, as an extension of the holographic model $\rho _{de}=3\alpha
H^{2}$ proposed by M. Li \cite{MLi}. \textit{\ }In the last case,
the key idea is to use the holographic principle \cite{Bousso} and
its possible role in  cosmology.\textit{\ } This approach is an
open issue and, under this philosophy, the model given in
(\ref{eq:eq25}) it is an interesting start point in order to
visualize what we mean by dark energy. This is a crucial fact if
we are thinking (the usual) in dark energy seen as a cosmological
constant, although the observational data would indicate that
$\rho _{de}$\ is not necessarily a constant \cite{Hu}. In this
sense, the\textit{\ }$ \Lambda CDM$ model could be questioned,
despite their successes.

By using (\ref{eq:eq25}) besides (\ref{eq:eq1}) and (\ref{eq:eq6}), it is possible to write
\begin{equation}
\rho _{de}\left( z\right) =3H^{2}\left( z\right) \left( \frac{2\alpha
-3\beta }{2+3\beta \omega _{de}\left( z\right) }\right) ,
\label{eq:eq26}
\end{equation}
and so
\begin{equation}
\rho _{dm}\left( z\right) =3H^{2}\left( z\right) \left(
\frac{2\left( 1-\alpha \right) +3\beta \left( 1+\omega _{de}\left(
z\right) \right) }{ 2+3\beta \omega _{de}\left( z\right) }\right)
, \label{eq:eq27}
\end{equation}
so that we can obtain an explicit expression for $Q$
\begin{equation}
\frac{Q}{9H^{3}}\left( z\right) =-\left( 2\alpha -3\beta \right)
\frac{\left[ 2\left( 1-\alpha \right) +3\beta \left( 1+\omega
_{de}\right) \right] \omega _{de}+\beta \left( 1+z\right) d\omega
_{de}/dz}{ \left( 2+3\beta \omega _{de}\right) ^{2}}.
\label{eq:eq28}
\end{equation}
The interaction sign can change through the evolution as can be
seen from the Ansatz $\omega _{de}\left( z\right) =\omega
_{de}\left( 0\right) +\sigma \;z$ (see later). Therefore,
according to (\ref{eq:eq28}) and by using the Ansatz given before,
$Q\left( z\right) $ experiences two sign changes, one in the past
and another in the future, as can be seen in \cite
{ArevaloCLP2014}. But, what does it mean a sign change of $Q$ at
late times?, would we have dominion of dark matter again?, the
current accelerated expansion would be transient previous to a
possible future collapse? \cite{KaloperPadilla2015}.

On the other hand, and under an entropic philosophy (entropic cosmology) in
which we have an amount of nonconservation energy, we can observe sign
changes in it and, in particular, that sign is mainly dependent of the
equation of state parameter ($\omega $) in each stage of the cosmic
evolution \cite{LepePena2014}. But, models based in entropic considerations
appear to be somewhat inconsistent with the observational data \cite{Gomez}.

And, under a holographic philosophy also, if we are considering an
interaction between the bulk and the boundary of the spacetime, we can see
sign changes in $Q\left( z\right) $ \cite{LepePenaTorres2015}.

Now, by using (\ref{eq:eq28}) besides (\ref{eq:eq27}) and
(\ref{eq:eq26}) into (\ref {eq:eq8}), we write
\begin{eqnarray}  \label{eq:eq29}
\omega _{de}^{eff}=\left( \frac{2\alpha -3\beta }{2+3\beta \omega _{de}}
\right) \left[ \omega _{de}-\frac{\beta \left( 1+z\right) }{2\alpha -3\beta }
\frac{d\omega _{de}}{dz}\right] ,
\end{eqnarray}
and
\begin{eqnarray}  \label{eq:eq30}
\omega _{dm}^{eff}=\omega _{dm}+\left( \frac{2\alpha -3\beta
}{2+3\beta \omega _{de}}\right)  \left[ \omega _{de}+\frac{\beta
\left( 1+z\right) }{ 2\left( 1-\alpha \right) +3\beta \left(
1+\omega _{de}\right) }\frac{d\omega _{de}}{dz}\right] ,
\end{eqnarray}
and $Q\left( z\right) $ can be written in the form
\begin{eqnarray}  \label{eq:eq31}
\frac{Q}{9H^{3}}\left( z\right) =-\frac{1}{3\beta ^{2}\sigma }\left( 2\alpha
-3\beta \right) \frac{\left( z-1.86\right) \left( z+0.2\right) }{\left(
A+z\right) ^{2}},
\end{eqnarray}
and the temperatures are given, respectively, by
\begin{eqnarray}  \label{eq:eq32}
T_{de}\left( z\right) =T_{de}\left( 0\right) \left( 1+z\right) ^{a}\left( 1+
\frac{z}{A}\right) ^{b},
\end{eqnarray}
and
\begin{eqnarray}  \label{eq:eq33}
T_{dm}\left( z\right) =T_{dm}\left( 0\right) \left( 1+z\right) ^{a}\left( 1+
\frac{z}{A}\right) ^{b+1}\left( 1+\frac{z}{B}\right) ^{c},
\end{eqnarray}
where
\begin{eqnarray}  \label{eq34}
a &=&\left( \frac{2\alpha -3\beta }{\beta \sigma }\right) \left( \sigma +
\frac{\omega \left( 0\right) -\sigma A}{A-1}\right) ,  \nonumber \\
b &=&-1-\left( \frac{2\alpha -3\beta }{\beta \sigma }\right) \left( \frac{
\omega \left( 0\right) -\sigma A}{A-1}\right) ,  \nonumber \\
c &=&2\alpha -3\beta ;  \nonumber \\
A &=&\frac{2+3\beta \omega \left( 0\right) }{3\beta \sigma }\text{ \ },\text{
\ }B=A-\left( \frac{2\alpha -3\beta }{3\beta \sigma }\right) ,
\end{eqnarray}
and we have considered $\omega _{dm}=0$ (dust). The parameters
involved are: $\omega _{de}(0)=-1.29$, $\sigma =0.47$, $\alpha
=0.73$ and $\beta =0.38$. All these parameters were estimated by
using an adjustment with type I Supernovae (Union 2)
\cite{ArevaloCLP2014, Kowalski2008} and from them we have $Q\left(
1.86\right) =Q(-0.2)=0$, and $a\approx 264.00$, $ b\approx
-264.15$, $c\approx 0.32$; $A\approx 0.98$ and $B\approx 0.39$. In
order to have a good vision of the Figures shown below, the
following limits can be obtained from
(\ref{eq:eq32}-\ref{eq:eq33})
\begin{eqnarray}  \label{eq35}
T_{de}\left( z\rightarrow -A\right) &\rightarrow &\infty \text{ \ \ },\text{
\ \ }T_{de}\left( z\rightarrow \infty \right) \rightarrow 0,  \nonumber \\
T_{dm}\left( z\rightarrow -B\right) &\rightarrow &0\text{\ \ },\text{ \ \ }
T_{dm}\left( z\rightarrow \infty \right) \rightarrow 0,
\end{eqnarray}
and
\begin{eqnarray}  \label{eq36}
\frac{T_{de}}{T_{dm}}\left( z\rightarrow -B\right) \rightarrow \infty \ \ ,\
\ \frac{T_{de}}{T_{dm}}\left( z\rightarrow \infty \right) \rightarrow 0.
\end{eqnarray}
According to (\ref{eq:eq29}-\ref{eq:eq30}), we note that if
$\omega _{de}=\mbox{const}.$ and $\omega _{dm}=0$ both
temperatures are equal. This last situation does not appears to be
consistent, is this a signal that $\omega _{de}\neq \mbox{const}.$
through the evolution?

\begin{figure}[!ht]
\begin{center}
\includegraphics[width=6cm]{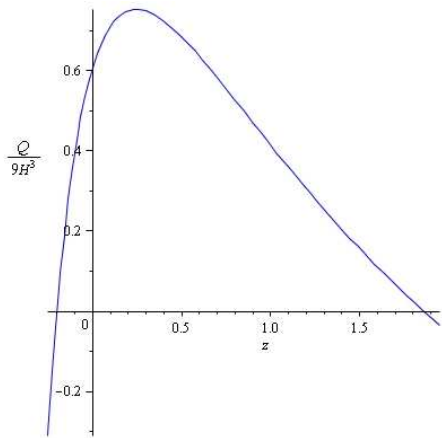}
\end{center}
\caption{The behaviour of $Q$ and its two sign changes: for $-0.2<z<1.86$ we
have $Q>0$, and $Q<0$ otherwise.}
\label{Figure1}
\end{figure}

\begin{figure}[!ht]
\begin{center}
\includegraphics[width=6cm]{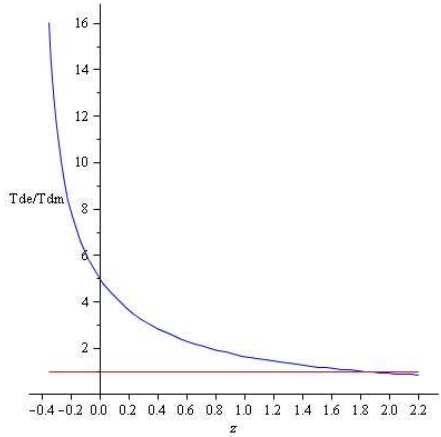}
\end{center}
\caption{We see that $T_{de}>T_{dm}$ in the range $-0.2<z<1.86$,
i.e., when $ Q>0$ and $T_{de}<T_{dm}$ for $z>1.86$, i.e., when
$Q<0$, in accord to the second law. However, in the range
$-0.39<z<-0.2$, when $Q<0$, the quotient $ T_{de}/T_{dm}$
diverges.} \label{Figure2}
\end{figure}

\begin{figure}[!ht]
\begin{center}
\includegraphics[width=6cm]{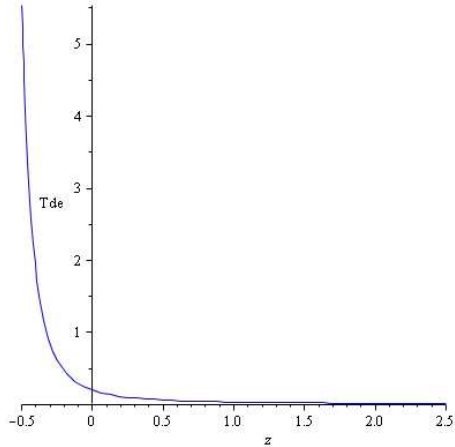}
\end{center}
\caption{We see that $T_{de}$ always grows and the dark energy exhibits a
negative heat capacity from $z\approx 1.86$ to $z\approx -0.2$ (dark energy
is heated while being delivered energy to dark matter). Out the indicated
range, we have a positive heat capacity. And $T_{de}$ diverges when $z$ goes
to $-0.98$.}
\label{Figure3}
\end{figure}

\begin{figure}[!ht]
\begin{center}
\includegraphics[width=6cm]{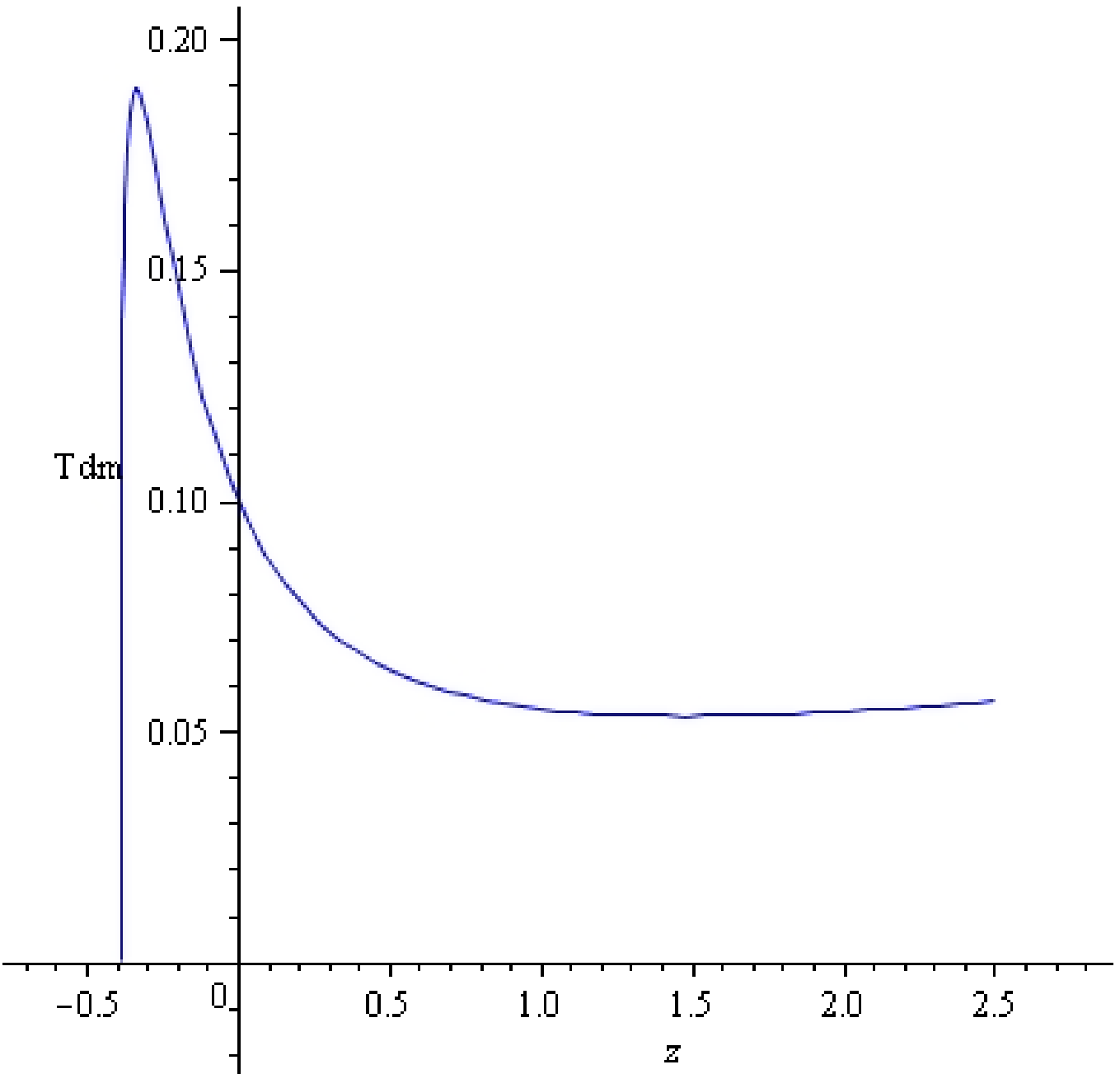}
\end{center}
\caption{We see that $T_{dm}\left( z\right) $ grows from $\infty $
to $ z=-0.33 $ (maximum of $T_{dm}$) and goes to zero when
$z=-0.39$. The dark matter heat capacity experiences a sign
change, that is, from $z=\infty $ to $z=1.86 $ we have a negative
heat capacity (dark matter is heated while being delivered energy
to dark energy), from $z=1.86$ to $z=-0.2$ the heat capacity is
positive, from $z=-0.2$ to $z=-0.33$ its heat capacity is negative
(dark matter cools while receiving energy) and from $z=-0.33$ to $
z=-0.38$ the heat capacity is positive.} \label{Figure4}
\end{figure}

So, sign changes in $Q$ imply changes in the temperatures of dark energy and
dark matter, as dictates the second law. Additionally, we can visualize
phase transitions, sign changes in its heat capacities, for both energy
densities. See Appendix for details.

Clearly, the observational data eventually will tell us if these changes will
definitely occur. Today, the observational data is only just showing signals
of the presence of $Q$, but as we stated above, appears unnatural to think
that cosmic fluids coexist and do not interact with each other.
Additionally, nothing we can say about values of $T_{de}\left( 0\right) $
and $T_{dm}\left( 0\right) $: future observations could elucidate this point.

And roughly speaking,\textbf{\ }the sign change in the cosmic
acceleration ($ z\sim 0.6$) is located "inside" the zone where
$Q\left( z\right) >0$.

\section{FINAL REMARKS}

Consistently with the second law of thermodynamics, we have studied the
behaviour of the temperatures of two interacting fluids (dark energy and
dark matter) and its relationship with the sign changes of $Q$ through the
cosmic evolution. We have investigated the phase transitions (sign changes
of its heat capacities) experienced for both, dark energy and dark matter.
We have used a holographic model for the dark energy and, as usual, we have
considered a presureless fluid (dust) for dark matter.\newline

Finally, the presence of $Q$ is a fact already confirmed by observations, the
validity of our results are full dependent from that in the sense of
possible changes in $Q$\ that can be observed in future observations. If
this is so, very interesting consequences we should have for the late
cosmology, in particular, if definitely we are heating or cooling in the
sense of ``dark'' (radiation background: we are cooling).

\section*{Acknowledgments}

This work was supported from PUCV-VRIEA Grant N$^{0}$
037.448/2015, Pontificia Universidad Cat\'{o}lica de
Valpara\'{\i}so (S. L.) and DIUFRO Grant N$^{0}$ DI14-0007 of
Direcci\'{o}n de Investigaci\'{o}n y Desarrollo, Universidad de La
Frontera (F. P.).
\appendix*

\section{Heat capacities}

\label{appendix} We write
\[
C=\frac{\Delta U}{\Delta T},
\]
and by doing $\Delta U=Q$ and recalling
\[
\dot{\rho}_{de}+3H\left( 1+\omega _{de}\right) \rho _{de}=-Q\text{ \ \ \ \ }
\mbox{and}\text{\ \ \ \ }\dot{\rho}_{dm}+3H\rho _{dm}=Q,
\]
(\ref{eq:eq6}) we write
\begin{eqnarray*}
C_{de} &=&\frac{-Q}{\Delta T_{de}}<0,\text{ \ }if\text{ \ }Q>0\text{ \ \ \ }
\mbox{and}\text{ \ \ }\Delta T_{de}>0, \\
\text{\ }C_{dm} &=&\frac{+Q}{\Delta T_{dm}}>0,\text{ \ }if\text{ \ }Q>0\text{
\ \ \ }\mbox{and}\text{ \ }\Delta T_{dm}>0.
\end{eqnarray*}

The temperatures. We see that $T_{de}\left( z\right) $ always grows with $z$
($\Delta T_{de}\left( z\right) >0$) in the range $-0.2<z<1.86$, see Fig. 3.
Then, the dark energy heat capacity is

\[
C_{de}\left( -0.2<z<1.86\right) =\frac{-Q}{\Delta T_{de}}<0,
\]
and out this range $C_{de}=+Q/\Delta T_{de}>0$.

We see also that $T_{dm}$ grows with $z$ from $\infty $ to $-0.33$ (maximum
of $T_{dm}$) and goes to zero when $z=-0.39$. In this range $\Delta
T_{dm}\left( z\right) >0$, but in the range $-0.39<z<-0.33$ we have $\Delta
T_{dm}\left( z\right) <0$. Then, the dark matter heat capacity changes are
\begin{eqnarray*}
C_{dm}\left( 1.86<z<\infty \right) =\frac{-Q}{\Delta T_{dm}}<0\;\;
\mbox{and}\;\;\Delta T_{dm}>0, \\
C_{dm}\left( -0.2<z<1.86\right) =\frac{+Q}{\Delta T_{dm}}>0\;\;\mbox{and}
\;\;\Delta T_{dm}>0, \\
C_{dm}\left( -0.2<z<-0.33\right) =\frac{-Q}{\Delta T_{dm}}<0\;\;\mbox{and}
\;\;\Delta T_{dm}>0, \\
C_{dm}\left( -0.33<z<-0.39\right) =\frac{-Q}{\Delta T_{dm}}>0\;\;
\mbox{and}\;\;\Delta T_{dm}<0.
\end{eqnarray*}

\end{document}